\documentstyle[11pt]{article}
\begin{document}
\title{  \sc In-plane magnetic field induced anisotropy\\ of 2D Fermi
contours and the field dependent cyclotron mass }

\author{ L. Smr\v{c}ka, T. Jungwirth \\
{\normalsize Institute of Physics, Academy of Science
of the Czech Republic,} \\
{\normalsize Cukrovarnick\'{a} 10, 162 00 Praha  6,
Czech Republic}}
\date{}
\maketitle
\vspace{3mm}

\noindent {\bf Abstract.} The electronic structure of a 2D gas
subjected to a tilted magnetic field, with a strong component
parallel to the GaAs/AlGaAs interface and a weak component
oriented perpendicularly, is studied theoretically. It is shown
that the parallel field component modifies the originally
circular shape of a Fermi contour while the perpendicular
component drive an electron by the Lorentz force along a Fermi
line with a cyclotron frequency given by its shape. The
corresponding cyclotron effective mass is calculated
self-consistently for several concentrations of 2D carriers as
a function of the in-plane magnetic field. The possibility to
detect its field-induced deviations from the zero field value
experimentally is discussed.

\vspace{8cm}
submitted to {\it Journal of Physics: Condensed Matter}

\thispagestyle{empty}
\newpage

\section*{1. Introduction}
The energy level structure of quasi-two-dimensional systems in
magnetic fields tilted with respect to the sample plane have
attracted attention for many years. As shown by Maan \cite{1} and
Merlin \cite{2} the electron spectra can be found analytically
for a special case of parabolic quantum wells. This simple
analytically solvable model is very useful when discussing
qualitative aspects of 2D electron physics in quantum wells, but
when the semi-quantitative comparison of results with
experimental data is the goal of investigation, more realistic
models must be used and numerical methods of calculation
employed.

The widely accepted approach to the electronic structure of
realistic quantum wells in tilted magnetic fields, which relies
on the perturbation theory, was reviewed by Bastard \cite{3}. It
can be outlined as follows. The Hamiltonian is decomposed into
three parts, $H = H_{\parallel} + H_{\perp} + H_{int}$. The first
part $H_{\parallel}$ describes the motion of an electron in
$z-$direction under the combined influence of the confining
potential and the in-plane component of the magnetic field. If
the in-plane field is not too strong its effect is usually
included via the perturbation theory. The second part of the
Hamiltonian $H_{\perp}$ corresponds to the standard 2D gas
subjected to the perpendicular component of the field and can be
diagonalized analytically. In most situations the third coupling
term $H_{int}$ can be completely neglected or treated as a small
perturbation. Only in certain special cases, e.g. when crossing
of Landau levels from different subbands is important, the matrix
elements of a coupling term must be evaluated and the Hamiltonian
$H$ diagonalized numerically \cite{4}.

In this paper we present a slightly different approach to the
electron structure of quantum wells in tilted magnetic fields
which is appropriate when the perpendicular component of the
applied magnetic field is weak. Assuming for a moment ${\bf B}
\equiv (0,B_y,0)$, i.e. the field exactly parallel to the plane
of 2D electron gas, the Hamiltonian $H_{\parallel}$ of an
electron confined to the $x-y$ plane by potential $V_{conf}(z)$
can be written as

\begin{equation}
H_{\parallel} = \frac{1}{2m}\left(p_x -e B_y z\right)^2 +
    \frac{1}{2m}p_y^2+\frac{1}{2m}p_z^2 +V_{conf}(z).
\end{equation}

\noindent Its energy spectrum is formed by subbands, momentum
operators $p_x$ and $p_y$ commute with $H_{\parallel}$ and the
eigenenergies $E_n(k_x,k_y), n = 0,1,\dots$ are the functions of
the quasi-continuous wavevectors $k_x$ and $k_y$. The 2D Fermi
surfaces, or more accurately the Fermi contours, are then
defined separately for each subband by the equation $E_F
=E_n(k_x,k_y)$.

The in-plane magnetic field $B_y$ does not influence the electron
motion in the $y-$direction while the electrons moving in the
$x-$direction are slowed down or accelerated by the combined
effect of the crossed fields $B_y$ and $E_z =
- dV_{conf}(z)/dz$, depending on the form of $V_{conf}(z)$.
Consequently, it is expected that the subband separation, the 2D
density of states and the shape of the Fermi contour will vary
with increasing $B_y$ \cite{3}. Only recently realistic
self-consistent calculations of electron energy spectra in
heterostructures subjected to parallel magnetic fields have been
performed \cite{5,6}, enabling the quantitative estimate of these
quantities.

Let the weak perpendicular component $B_z$ of the magnetic field
be added to the strong in-plane field component $B_y$. Then the
electron dynamics in the $x-y$ plane can be described
semiclassically, similarly as the dynamics of electrons in metals
with non-spherical Fermi surfaces. The corresponding
quasi-classical theory was originally developed by Onsager
\cite{7} and Lifshitz \cite{8}, and nowadays it is a part of
standard textbooks devoted to the solid state physics (see e.g.
\cite{9}). According to this theory electrons move in the
$k$-space along trajectories defined by intersections of a Fermi
surface and planes perpendicular to the applied magnetic field.
In the 2D case there is only one such line, identical with the
Fermi line itself. Once the $k$-space orbit is known, the
semiclassical theory predicts that the electron real space
trajectory is of an identical shape except of the scale factor
$\hbar/|e|B_z$ , and rotation by $\pi/2$. The anisotropy of the
Fermi contour due to the strong parallel field thus leads to the
deviation of the electron trajectory from the originally circular
form.

The modification of the the Fermi contour shape by the in-plane
magnetic field should manifest itself through the field
dependence of the cyclotron effective mass $m_c$. This important
characteristics of electron energy spectra is related to the
Fermi area $S_F$ surrounded by the Fermi contour by
a semi-classical expression

\begin{equation}
m_c = \frac{\hbar^2}{2\pi}\;\frac{dS_F}{dE}.
\end{equation}

\noindent Note that the field dependence of $m_c$ is determined
by the combined influence of the in-plane component of the
magnetic field and of the shape of the confining potential
$V_{conf}(z)$ of the quantum well and, therefore, the different
structures with different quantum wells can be distinguished by
measuring this quantity.

Two methods are widely used to determine the cyclotron effective
mass of 2D electrons confined to GaAs/AlGaAs interface: the
cyclotron resonance in the infrared region of optical spectra
\cite{10} and the temperature damping of Shubnikov - de Haas
oscillations \cite{11}. To get insight into the feasibility of
such experiments, we report in this paper the results of
self-consistent calculations of the field dependence of the
cyclotron effective mass in the electron layer confined to the
interface of the standard GaAs/AlGaAs heterostructure for several
concentrations of 2D carriers.

\section*{2. Electronic structure in tilted magnetic fields}

As mentioned above, non-interacting electrons mobile in the
$x-y$ plane and confined by the potential $V_{conf}(z)$ in the
$z$-direction are considered. A tilted magnetic field ${\bf B}
= (0,B_y,B_z)$ is applied to the system. The corresponding
one-electron Hamiltonian reads

\begin{equation}
 H = \frac{1}{2m}\left({\bf p} - e{\bf A}\right)^2 +
 V_{conf}(z)
\end{equation}

\noindent where $e$ and $m$ are electron charge and effective
mass, respectively. We choose the vector potential ${\bf A}$ in
the form ${\bf A} =(-B_z y+B_y z,0,0)$ which can be split into
a sum ${\bf A = A_{\perp}+A_{\parallel}}$ of vector potentials
${\bf A_{\perp}} =(-B_{z}y,0,0)$ and ${\bf A_{\parallel}}
=(B_{y}z,0,0)$ describing the perpendicular and parallel
components of the magnetic field, respectively.

To bring the above Hamiltonian to a form similar to (1) we
introduce new canonical momenta by

\begin{eqnarray}
\pi_x = p_x - e A_{\perp x} = p_x -m\omega_z y ,\nonumber \\
\\
\pi_y = p_y - e A_{\perp y} = p_y \hspace{1.2cm}   \nonumber
\end{eqnarray}

\noindent where $\omega_z$ denotes $|e|B_z /m$. These momenta
correspond to the in-plane electron motion and the perpendicular
component of the magnetic field $B_z$ enters Hamiltonian
exclusively through them.

While the quantum mechanics is inevitable to obtain the correct
picture of an electron motion in the $z-$direction, both quantum
mechanical and classical descriptions are acceptable for the
in-plane motion when the perpendicular component of the magnetic
field $B_z$ is weak. The reason is that the weak perpendicular
component yields many occupied Landau levels below the Fermi
energy. The Landau levels near the Fermi energy are represented
by the states with high quantum numbers and, as it is well known,
they can be treated quasi-classically as well as
quantum-mechanically. Therefore, in the first step of our
approximate description the in-plane canonical variables and
momenta (including the part ${\bf A_{\perp}}$ of the vector
potential) will be considered as classical variables which
commute with the Hamiltonian and enter it as $c-$numbers.
Introducing further the components of the  wavevector, $k_x$ and
$k_y$, by $\pi_x = \hbar k_x$ and $\pi_x = \hbar k_x$,
respectively, we obtained the one-dimensional Hamiltonian
identical to (1).

Again, the energy spectrum of this Hamiltonian is formed by
subbands and the eigenenergies $E_n(k_x,k_y)$ are the continuous
functions of the wavevectors $k_x$ and $k_y$ or, equivalently,
of the in-plane canonical momenta $\pi_x$ and $\pi_y$. Since the
perpendicular component of the magnetic field is hidden in the
wave vector component $k_x$, the Hamiltonian describes the 2D
system subjected only to the parallel field ${\bf B} = (0, B_y,
0)$. The shape of subbands is determined by both the confining
potential and the in-plane magnetic field and can be accurately
obtained by the self-consistent numerical calculation.

In the second step of our approximate treatment the
single-subband effective Hamiltonian $H_{eff}$ for the in-plane
electron motion is constructed from a subband energy
$E_n(k_x,k_y)$ by the backward substitution $k_x \rightarrow
\pi_x/\hbar$, $k_y \rightarrow \pi_y/\hbar$. We get

\begin{equation}
H_{eff} = E_n(\pi_x, \pi_y)
\end{equation}

\noindent and this effective Hamiltonian will be used to describe
the electron motion in the $x - y$ plane by standard Hamilton
equations $\dot{x} = \partial H_{eff}/\partial p_x$,
$\dot{y} = \partial H_{eff}/\partial p_y$,
$\dot{p_x} = - \partial H_{eff}/\partial x$,
$\dot{p_y} = - \partial H_{eff}/\partial y $.
The time derivatives of  coordinates  define the
velocity components $v_x = \dot{x}, v_y = \dot{y}$
and yield expressions

\begin{equation}
v_x = \frac{1}{\hbar}\,\frac{\partial E_n}{\partial k_x} , \;\;\;
v_y = \frac{1}{\hbar}\,\frac{\partial E_n}{\partial k_y}.
\end{equation}

\noindent
The time derivatives of momenta lead to

\begin{equation}
\hbar \dot{k_x} = - |e|B_z\,v_y  , \;\;\;
\hbar \dot{k_y} =  |e|B_z\,v_x .
\end{equation}

\noindent These are the classical equations of motion for an
electron subjected to the Lorentz force $ e\,({\bf v \times B})$,
${\bf B} = (0, 0, B_z)$. Their solutions are in the form of
orbits of all energies, among them the orbit corresponding to the
Fermi energy is the most important. A simple geometrical analysis
allows to derive the following properties of orbits.

As already mentioned, an electron is driven by the Lorentz force
around the Fermi contour in the $k$-plane. In the $x-y$ plane, it
executes an orbit similar in shape but scaled in dimensions by
$\hbar/|e|B_z$ and turned through $\pi/2$. It has also a
$z$-component of motion which will be discussed later.

The period $T$ of the cyclotron motion, or equivalently the
cyclotron frequency $\omega_c$, are obtained from (6,7) by the
direct integration. Since the magnitude of these parameters
depend linearly on the amplitude of the perpendicular field
component, it is more convenient to characterize the orbits by
the cyclotron effective mass $m_c$ defined with a help of the
equation $\omega_c =|e|B_z/m_c$. The explicite expression
relating $m_c $ to the shape of the Fermi contour reads

\begin{equation}
m_c = \frac{\hbar^2}{2\pi}\;\oint \frac{dk}{|\nabla_kE|}
\end{equation}

\noindent where $dk$ denotes an element of a length of the Fermi
line. More often the equivalent expression (2) is used.

It is well known that, unlike in the 3D case, the area $S_F$
surrounded by the $k$-space orbit is in 2D systems identical to
the Fermi area which is determined by the concentration $N_e$ of
free carriers. Since the density of states $g$ is related to the
concentration by $g = dN_e/dE$ we can write

\begin{equation}
N_e =  \frac{2}{(2\pi)^2}\;S_F,\;\;\;\; g = \frac{2}{(2\pi)^2}\;
\frac{dS_F}{dE},
\end{equation}

\noindent where a spin degeneracy is included. Making use of these
expressions and of the equation (2) a simple relation between the
density of states and the cyclotron mass is obtained

\begin{equation}
g = \frac{m_c}{\pi \hbar^2}.
\end{equation}

The quantization of the in-plane electron motion by $B_z$ can be
taken into account using the Bohr-Sommerfeld quasi-classical
quantization rules which state that each quantized real space
trajectory encloses an integer number of flux quanta $\hbar
/|e|$. This procedure yields a discrete spectrum of Landau levels
and the density of states becomes a series of delta functions
separated by $\hbar\omega_c$. Note that the filling factor of
each level is $2|e|B_z/h$, i.e. independent of the cyclotron mass
and the same as in the case of exactly perpendicular magnetic
field.

To estimate the validity of our approximate treatment of the
electron spectra we consider a model in which a 2D system is
confined to the $x-y$ plane by a harmonic potential $V_{conf}(z)
= m\,\Omega^2z^2/2$. In this special case the eigenenergies can
be found in an analytic form both for in-plane and tilted
magnetic fields \cite{1}. Assuming first, as before, $B_z = 0$ we
can write

\begin{equation}
E_n(k_x,k_y) = \hbar\widetilde{\omega}(n+\frac{1}{2})
              + \frac{\hbar^2k_x^2}{2\widetilde{m}}
              + \frac{\hbar^2k_y^2}{2m},\;\;\;\;
               n = 0,1,\dots
\end{equation}

\noindent  where  $\widetilde{\omega}  =(\omega_y^2  +  \Omega^2)
^{1/2}$, $\;\omega_y = |e|B_y/m$ and $\widetilde{m} = m\,
\widetilde{\omega}^2/\Omega^2$.

The magnetic field induced anisotropy is described by the new
effective mass $\widetilde{m}$ for the $x$-component of the
electron motion which grows up with $B_y$. The Fermi contour is
distorted from the circular shape to the eliptic one. The
zero-field Fermi radius $k_{0F}$ is determined by $E_{0F} =
\hbar^2 k^2_{0F}/2m$ where $E_{0F}$ is the zero-field Fermi
energy. Taking into account the changes in the energy spectrum
and the Fermi energy due to the in-plane magnetic field, the
equation $E_F = E_n(k_x,k_y)$ for determination of the Fermi
contour can be rewritten in the form

\begin{equation}
k^2_{0F} = \frac{\Omega}{\widetilde{\omega}}\,k^2_x
          +\frac{\widetilde{\omega}}{\Omega}\,k^2_y.
\end{equation}

\noindent From this equation the cyclotron effective mass can be
evaluated using (8) and we get

\begin{equation}
m_c = \sqrt{m \widetilde{m}}.
\end{equation}

\noindent The corresponding cyclotron frequency can be written
as $\omega_c = \Omega\omega_z/\widetilde{\omega}$.

This approximate results should be compared with the exact
solutions obtained for the parabolic quantum well subjected to
tilted magnetic fields:

\begin{equation}
\omega_{1,2} = \frac{ \sqrt{\omega_y^2 + (\omega_z + \Omega)^2}
             \mp \sqrt{\omega_y^2 + (\omega_z - \Omega)^2}}{2}.
\end{equation}

\noindent It is easy to show that the lowest order expansion of
these expressions with respect to $\omega_z$ yields exactly the
results obtained by our approximate method

\begin{equation}
\omega_1 \approx \frac{\Omega}{\widetilde{\omega}}\,\omega_z,
\;\;\;\;
\omega_2 \approx \widetilde{\omega}.
\end{equation}

\noindent Both exact and approximate solutions are shown in
figure 1.

%%%%%%%%%%%%%%%%%%%%%%%%%%%%%%%%%%%%%%%%
\section*{3. Self-consistent calculations}

The standard semi-empirical model working quantitatively for the
lowest conduction states of GaAs/AlGaAs heterostructures is used
to solve the Schr\"{o}dinger equation in the envelope function
approximation. The envelope function is assumed to be built from
host quantum states belonging to a single parabolic band. Since
the effect of the effective mass mismatch is completely neglected
and the envelope functions of GaAs and AlGaAs are smoothly
matched at the interface, the Schr\"{o}dinger equation has a form
given by (1).

The confining potential

\begin{equation}
V_{conf}(z)=V_b(z)+V_{s.c.}(z)
\end{equation}

\noindent is a sum of the step function $V_b(z)=V_b\,\theta(-z)$
corresponding to the conduction band discontinuity between AlGaAs
and GaAs and of a term describing the interaction of an electron
with ions and the electron-electron interaction. This term should
be calculated self-consistently and can be written as

\begin{equation}
V_{s.c.}(z)=V_H(z)+V_{xc}(z).
\end{equation}

\noindent The Hartree term $V_H$ is determined from the Poisson
equation

\begin{equation}
\frac{d^{2}V_{H}}{dz^{2}}=\frac{|e|\varrho(z)}{\varepsilon}
\end{equation}

\noindent and  we use
an expression calculated by Ruden and D\"{o}hler \cite{12} in
a density-functional formalism for the exchange-correlation term
$V_{xc}$:

\begin{equation}
V_{xc}\simeq-0.611\frac{e^2}{4\pi\varepsilon}
\left(\frac{3N_e(z)}{4\pi}\right)^{1/3}.
\end{equation}

\noindent The conduction band offset $V_b$ and the dielectric
constant $\varepsilon$ enter our calculations as input
parameters.

For modulation doped GaAs/AlGaAs heterostructures the total
charge density $\varrho(z)$ in the equation (18) can be split
into parts corresponding to concentrations of electrons,
$N_e(z)$, their parent donors in AlGaAs, $N_d^+(z)$, and ionized
residual acceptors in GaAs, $N_a^-(z)$:

\begin{equation}
\varrho(z)=e\left[N_e(z)-N_d^+(z)+N_a^-(z)\right].
\end{equation}

\noindent We accept a usual approximation of constant impurity
concentrations and assume donors and acceptors to be ionized
within certain finite intervals $l_d$ and $l_a$: $N_d^+(z)=N_d$
for $-l_d-w\leq z\leq -w$ and $N_a^-(z)=N_a$ for $0\leq z \leq
l_a$, $w$ is the spacer thickness.

In our calculation, we consider a GaAs/AlGaAs heterostructure
with parameters $N_d=2\times 10^{18}$ $\rm cm^{-3}$,
$N_a=10^{14}$ $\rm cm^{-3}$, the band offset $V_b=225$ $\rm meV$ and
the dielectric constant $\varepsilon = 12.9$. Three selected
values of the spacer thickness $w_1 = 40$ $\rm nm$, $w_2 = 20$ $\rm
nm$ and $w_3 = 10$ $\rm nm$ yield electron systems with three
different concentrations of electrons
$N_{e1}\approx 1.8\times 10^{11}$ $\rm cm^{-2}$,
$N_{e2}\approx 3.4\times 10^{11}$ $\rm cm^{-2}$ and
$N_{e3}\approx 5.4\times 10^{11}$ $\rm cm^{-2}$, respectively. All
these systems have only one occupied subband. The parameters
$l_a$ and $l_d$ are determined in the course of the
self-consistency procedure. For more details see [6].

The \lq egg-like\rq{} Fermi contours calculated for the above
three concentration and several magnitudes of the in-plane
magnetic fields are shown in figure 2. The real space
trajectories have similar shapes. Let us note that, since the
average value of the out-of-plane coordinate of an electron
$\langle z \rangle_{k_x}$ is a function of $k_x$, i.e. the
position on the Fermi contour , the resulting trajectory does not
lie exactly in the $x - y$ plane but is slightly tilted. It is
elongated in the $y$ direction and an electron is close to the
interface at wide end of the trajectory and in the bulk at its
narrow end. A schematics of a classical
real space
trajectory illustrating this behaviour is shown, together with
its projection to the $x - y$ plane, in figure 3.

The field dependence of the cyclotron effective mass resulting
from the self-consistent calculations is presented in figure 4.
For all three electron concentrations the deviations of the
cyclotron effective mass reach almost 25\% of its zero field
value in magnetic field $B_y = 10$ T.

\section*{4. Discussion and summary}

The electronic structure of a 2D electron gas in GaAs/AlGaAs
heterostructure subjected to a tilted magnetic field has been
studied theoretically. We combine (i) the self-consistent quantum
mechanical calculations of the electron subbands of the 2D
electron system in the presence of the parallel magnetic field
and (ii) the subsequent quasi-classical description of the
in-plane electron motion under the influence of the perpendicular
magnetic field component.

The parallel magnetic field combined with the confining effect of
the quantum well is the reason for deviations of the Fermi
contour from a circular shape. Note that in this case the nature of changes
is different than in case of a standard modification of the
Fermi line due to the
periodic potential. Here, the time-reversal
symmetry is broken by the magnetic field and,
consequently, the Fermi contour
has no inversion symmetry in the $k$ space, if the
quantum well is asymmetric.

The quasi-classical quantization of the in-plane electron motion
yields the eigenenergies linear in the perpendicular component of
the magnetic field. The method is applicable if the energy
separation of subbands is larger than the separation of the
Landau levels. It does not necessarily mean that
the parallel magnetic field
component must be always greater than the perpendicular one,
but the in-plane field increases the subband
separation.

The  deviations of eigenenergies from the linear
dependence for higher perpendicular magnetic fields are
attributed to the fact that the real space trajectories do not
lie exactly in the $x - y$ plane, as already mentioned in
previous sections. In this case the quantization rules should not take
into account only the perpendicular
component of the magnetic field and the area surrounded by the
projection of a trajectory to the $x - y$ plane but the full
field and the area of the trajectory itself.

The in-plane magnetic field dependencies of the cyclotron mass
were calculated in the limit of the infinitesimally small perpendicular
component of the magnetic field.
The changes of the cyclotron mass are surprisingly large and, to
our knowledge, fully in the scope of the present experimental
technique.

\section*{\bf Acknowledgments}

This work has been supported in part by the Academy of Science of
the Czech Republic under Contract No 11 059, and by the Ministry
of Education, Czech Republic under Contract No V091.

\newpage
\thispagestyle{empty}

\newpage
\thispagestyle{empty}
\noindent{\Large \bf Figure captions}\\

\noindent
{\bf Figure 1.} Eigenfrequencies of a parabolic quantum well subjected
to a tilted magnetic field, $\varphi=60^o$. Dashed lines denote the exact
solution, full lines the approximate one.\\

\noindent
{\bf Figure 2.} Self-consistently calculated Fermi lines corresponding
to electron concentrations $N_{e1}\approx 1.8\times 10^{11}$ $\rm cm^{-2}$,
$N_{e2}\approx 3.4\times 10^{11}$ $\rm cm^{-2}$,
$N_{e3}\approx 5.4\times 10^{11}$ $\rm cm^{-2}$ for: (a) B=0 T, (b) B=5 T,
(c) B=10 T.\\

\noindent
{\bf Figure 3.} Schematics of a classical real space trajectory of an
electron in an electric and a tilted magnetic field (thick line) and
its projection to the $x-y$ plane (thin line).\\

\noindent
{\bf Figure 4.} Self-consistently calculated in-plane magnetic field
dependence of the relative cyclotron mass corresponding to electron
concentrations $N_{e1}\approx 1.8\times 10^{11}$ $\rm cm^{-2}$,
$N_{e2}\approx 3.4\times 10^{11}$ $\rm cm^{-2}$,
$N_{e3}\approx 5.4\times 10^{11}$ $\rm cm^{-2}$.

\begin{thebibliography}{10}


\bibitem{1} Maan J C 1984,{\it Two-Dimensional Systems,
Heterostructures and Superlattices} ed G Bauer et al (Berlin:
Springer) p 183

\bibitem{2} Merlin R 1987 {\it Solid State Commun.} {\bf 64} 99

\bibitem{3} Bastard G 1990 {\it Wave mechanics applied to
semiconductor heterostructures} les \'{e}ditions de physique
(Paris) p 317

\bibitem{4} Lee S J, Park M J, Ihm G, Falk M L, Noh S K, Kim
T W and Choe B D 1993 {\it Physica B} {\bf 184} 318

\bibitem{5} Heisz J M and Zaremba E 1993 {\it Semicond. Sci
Technol.} {\bf 8} 575

\bibitem{6} Jungwirth T and Smr\v{c}ka L 1993 {\it J. Phys. C.:
Condens. Matter} {\bf 5} L 217

\bibitem{7} Onsager L 1952 {\it Phil. Mag.} {\bf 43} 1006

\bibitem{8} Lifshitz I M 1956 {\it Sov. Phys.- JETP} {\bf 30} 63

\bibitem{9} Ashcroft N W and Mermin N D 1976 {\it Solid State
Physics} (Philadelphia: Saunders College)

\bibitem{10} Warburton R J, Watts M, Nicholas R J Harris and
Foxon C J 1992 {\it Semicond. Sci. Technol.} {\bf 7} 787

\bibitem{11} de Lange W, Blom F A P and Wolter J H 1993 {\it
Semicond. Sci. Technol.} {\bf 8} 341

\bibitem{12} Ruden P and D\"{o}hler 1983 G H {\it Phys. Rev. B}
{\bf27} 3538

\end{thebibliography}
\end{document}